\RequirePackage[2020-02-02]{latexrelease}
\providecommand\DeclareCommandCopy[2]{}
\documentclass[twocolumn,prl,amsmath,amssymb,showpacs,superscriptaddress,floatfix]{revtex4}

\usepackage{graphicx}
\usepackage{bm}
\usepackage{lineno,hyperref}
\usepackage{epsf}
\usepackage{xcolor}
\begin{document}
\title{Voids in liquids: peculiarities of molecular dynamics simulation of fluid systems}

\author{Yu. D. Fomin \footnote{Corresponding author: fomin314@mail.ru}}
\affiliation{Vereshchagin Institute of High Pressure Physics,
Russian Academy of Sciences, Kaluzhskoe shosse, 14, Troitsk,
Moscow, 108840, Russia }
%

%

%
\date{\today}

%

\begin{abstract}
Molecular dynamics is a powerful tool to investigate the properties of fluid systems.
However, a correct interpretation of the results of simulations is required. In particular,
some simulations show appearance of large voids in liquids, which contradicts our
common sense on what is liquid. In the present paper we discuss the origin of large cavities
liquids in molecular dynamics simulations. We demonstrate that the cavities appear either
if the temperature of the system is above the critical temperature of liquid-gas transition
or if the system is in two-phase liquid-gas region. These conclusions are illustrated by several
examples from literature and our own simulations.

{\bf Keywords}:molecular dynamics simulation, supercritical fluid, two-phase region, void, cavity
\end{abstract}

\pacs{61.20.Gy, 61.20.Ne, 64.60.Kw}

\maketitle


\section{Introduction}

Molecular dynamics (MD) is an important computational tool which often give the results which can effectively substitute the experimental ones.
At the same time molecular dynamics simulations are usually faster, simpler and cheaper than real experiments. It makes MD
an important tool to study the properties of substances.

MD is of particular importance in investigation of liquids. The data from MD allows to trace trajectories of every
particle, which allows to calculate the properties, which are difficult (but not impossible) to determine in experiments:
structural properties of the liquid, diffusion coefficient, volume of the Voronoi cells of the particles, etc.

We are not going to describe all possible applications of molecular dynamics to investigation of liquids: this topic is to
widespread to give even a brief description. In the present manuscript we would like to point out a paradoxical observation
obtained in MD simulations: it was found that some liquids demonstrate formation of voids (a term 'cavity' is used in some
publications). From our everyday experience we know that voids can appear in solids, but not in liquids. However,
there are many publications where nanovoids, i.e., microscopic cavities big comparing to the volume of Voronoi cells
of the particles are observed. We would like to emphasize that we consider only big voids, which is in contrast to small
voids considered in solutions (these voids are of the size comparable to the molecular one, which allows to introduce an
allien molecule in the given solvent). A theory of such small voids in hard sphere system was constructed by
Debenedetti and Truskett in Ref. \cite{deb}. The authors define the free volume as the volume where a given sphere
can move if all other spheres are fixed. They construct the probability distribution of the free volume of hard spheres
at the density $\rho*\sigma^3=0.97$ ($\sigma$ is the diameter of the spheres) and find that the free volume does not
exceed $\sigma^3$ (see Fig. 3 of Ref. \cite{deb}), i.e., the cavities do not exceed the size of the spheres themselves.

At the same time much larger voids are reported in other liquids. One of examples of such liquids is liquid tellurium.
Combened X-ray and ab-initio molecular dynamics study of liquid tellurium is given by Akola et.al. in Ref. \cite{te}. The authors
find reasonable agreement between the experimental and computed structure factors which justifies the computational
procedure. Within the framework of ab-initio molecular dynamics they characterize the microscopic structure of
liquid tellurium and find large cavities (see Fig. 5(b) of Ref. \cite{te} for visualization of the cavities). They
also compute the size of the cavities which appears to be about $80$ $\AA^3$, which is much larger than the
Voronoi cell volume of the particles. The total volume of the cavities in liquid tellurium, according to
Ref. \cite{te}, is $26-35 \%$, i.e., about from a quater to one third of the total volume of the system.
Apparently, such large voids have another nature than the ones in hard spheres. The authors note that
cavities play important role in thermodynamic properties of tellurium, but do not explain their origin.

Such large voids are observed in other liquids too. Logunov and Orekhov report formation of cavities
in liquid carbon modelled with GAP-20 model \cite{or-gap}. The authors perform pore size distribution
calculations of carbon at $T=6000$ K and pressure from $0.5$ to $4.0$ GPa assuming that the diameter of the
atoms is $1.95$ $\AA$ and found that the most probably linear size of the pore varies from about 3 to 8 $\AA$ depending
on the pressure. At low pressure they observe the voids as large as about 10 $\AA$ in diameter.

Another case where voids in liquids is observed involves metastable states, for instance, superheated liquid
(see, for instance, Ref. \cite{voids-cu} for metastable copper and Ref. \cite{voids-al} for voids in aluminium
in the case of shock waves propagation). In the present study we limit ourselves to thermodynamically equilibrium phases.

As is seen from the discussion above large voids can be observed in different liquids. At the same time we are
not aware of any explanation of the origin of these voids. Importantly, the voids are observed in molecular dynamics simulations
only. Although it can be related to the difficulty of void detection in experiments, it might be also an
artefact of the simulation methodology.

The goal of the present study is to simulate several liquids with large voids by means of molecular dynamics method
and to find out the origin of the voids. Based on the understanding of the reasons of void formation we can predict the
thermodynamic conditions where voids in molecular dynamics can be observed.

\section{System and Methods}

Several systems were studied in the present work. First of all, we simulated classical Lennard-Jones (LJ) fluid:

\begin{equation}
U(r)=\varepsilon \left( \left( \frac{\sigma}{r} \right)^{12} - \left( \frac{\sigma}{r} \right)^{6} \right).
\end{equation}
The parameters $\varepsilon$ and $\sigma$ are used as the units of energy and length respectively and all quantities for LJ systems
are expressed in reduced units. The cut-off radius of LJ system was set to $r_c=2.5$. No tail correction were employed.

Two-dimensional (2d) LJ fluid is simulated first. A system of 20000 particles in a rectangular box was simulated. The temperature
is $T=0.5$ which is slightly above the critical one $T_c=0.49$ \cite{widom-2d}. The system was simulated for $5 \cdot 10^6$ steps. The last
$5 \cdot 10^5$ steps were taken for analysis. The timestep was taken $dt=0.001$.

Next we simulate supercritical carbon. GAP-20 potential \cite{gap-20} was used. The system of 8000 in a cubic box was simulated.
The densities varied from $0.05$ $g/ml$ to $2.0$ $g/ml$ with the step $\Delta \rho =0.05$ $g/ml$. The temperatures varied
from $T=4000$ K to $5000$ K with step $\Delta T=100$ K. A detailed description of the simulation methodology is described
in Ref. \cite{carbon-widom}.

In the case of three dimensional LJ fluid systems of two sizes were simulated: a small one (4000 particles) and
a large one ($1 \cdot 10^6$ particles). In both cases cubic boxes with periodic boundary conditions were used.
The densities varied from $\rho=0.1$ to $0.8$ with step $\Delta \rho=0.1$. Two temperatures were considered for
the small system: $T=0.5$ which is below the triple point temperature ($T_{tr}=0.694$ \cite{lj-tr}) and
$T=1.0$ which is in between the triple and critical temperature ($T_{c}=1.31$ \cite{lj-cr}). Only $T=1.0$
was simulated for the large system. The timestep was set to $dt=0.001$. $10 \cdot 10^6$ steps were performed
for small system and $2 \cdot 10^6$ for the large one.

Simulations of water were performed with a well recognized TIP4P/2005 model \cite{tip4p}. This model rather accurately
reproduces the boiling curve of water including critical temperature and critical density of water \cite{vega-vle},
which is principal for the present study. A system of 4000 water molecules in a cubic box with periodic boundary
conditions was used. The simulations were performed at two temperatures: $T=400$ and $570$ K. The pressure was
set to zero in the former case and three values of pressure ($P=0$, $55.43$ and $100$ bar) in the latter. The
value $P=55.43$ bar corresponds to the coexistence pressure at this temperature. The system was simulated both
in NVT and NPT ensembles. In the case of canonical ensemble the densities varied from $0.1$ $g/ml$ to
$1.2$ $g/ml$. In the case of isothermal-isobaric ensemble the initial configurations correspond to
the same set of densities. The results for the equation of state obtained in NVT and NPT ensemles
are compared.

All simulations described above were performed using the LAMMPS simulation package \cite{lammps}.

Finally liquid tellurium was simulated. Ab-initio molecular dynamics was performed with the help of Vienna Ab initio Simulation Package (VASP)
\cite{vasp1,vasp2,vasp3,vasp4}. Projector augmented wave (PAW) method was used to describe the electron-ion interaction
\cite{paw}. The PAW pseudopotential (dated 22Mar2012) with 6 valence electrons ($5s^2 5p^4$) and the
local density approximation (LDA) for exchange-correlcation functional were used \cite{lda}.

The system of 100 atoms in a rectangular box with periodic boundary conditions
was used. Following Ref. \cite{te} we firstly equilibrate the system at very high temperature $T=3000$ K for
5 ps. Then we use the obtained structure as an initial one for the simulations at lower temperatures. These simulations
also take 5 ps. The timestep is 1 fs. The pressure is set to zero. The box is allowed to change
its dimension in all directions, while the angles of the box are fixed. The energy cut-off is taken to 250 eV (the Enmax
in POTCAR file is 175.144 eV). Although it looks rather low, it is consistent with the Ref. \cite{te} which is used for comparison of the data.
Only $\Gamma$ point was used in Brillouin zone.

The visualization of the results was performed in OVITO software \cite{ovito}.

\section{Results and Discussion}

First of all we would like to stress that no large voids are possible in liquids. If large voids are observed, it means that the system is
not a liquid. Two situations are possible, which we are going to consider below: supercritical fluid and liquid-gas two-phase
regions.

\subsection{Supercritical fluid}

As the first situation we describe the case of a fluid at temperatures exceeding the critical temperature of the
gas-liquid phase transition. For the sake of brevity we will call these systems as "supercritical fluids" (SCF).
Indeed, there is no common agreement in the literature what is SCF. The most common definition also states,
that both temperature and pressure should exceed the critical one \cite{deben}, which is not the case of the present discussion. Some
studies involve the concepts of Widom \cite{widom-init,widom-lj} and Frenkel \cite{fr-1,fr-2} lines to determine the SCF. In the present study we use
this term for fluids at the temperature above $T_c$.

It is well known that at $T>T_c$ fluids can demonstrate numerous anomalous features, such as maxima
of thermodynamic properties \cite{widom-init,widom-lj,widom-sw,widom-co2}, clusterization \cite{widom-2025},
divergence of correlation length \cite{widom-sw}, etc. In the case of the present
work we are interesting in the clusterization properties of the SCF. It is well known that SCF has inhomogeneous structure:
at low densities clusters of several particles are located in gas-like media, while at higher ones SCF looks
like a liquid with voids (see, for instance, \cite{sedunov}). Although clustering structure of SCF cannot be probed
by usual methods of structure monitoring, like X-ray scattering, it can be experimentally verified by optical methods
\cite{mareev,mareev1}. At the same time the structure of SCF can be easily visualized by molecular simulation methods.

As the first example we consider a 2d LJ system. Such choice allows us to give very clear visualization of the
discussed features. After that we will turn to three dimensional systems.

The boiling line and lines of near-critical maxima of 2d LJ system were reported in our recent article
\cite{widom-2d}. The critical point of this system is $T_c=0.49$, $P_c=0.02858$ and $\rho_c=0.36$. We also
performed analysis of clusterization of the system and found the percolation threshold. In the case of
2d LJ system the density at which a percolation cluster appears does not depend on the temperature and is very
close to the critical density: $\rho_{p}=0.325$.

\begin{figure}

\includegraphics[width=9cm, height=7cm]{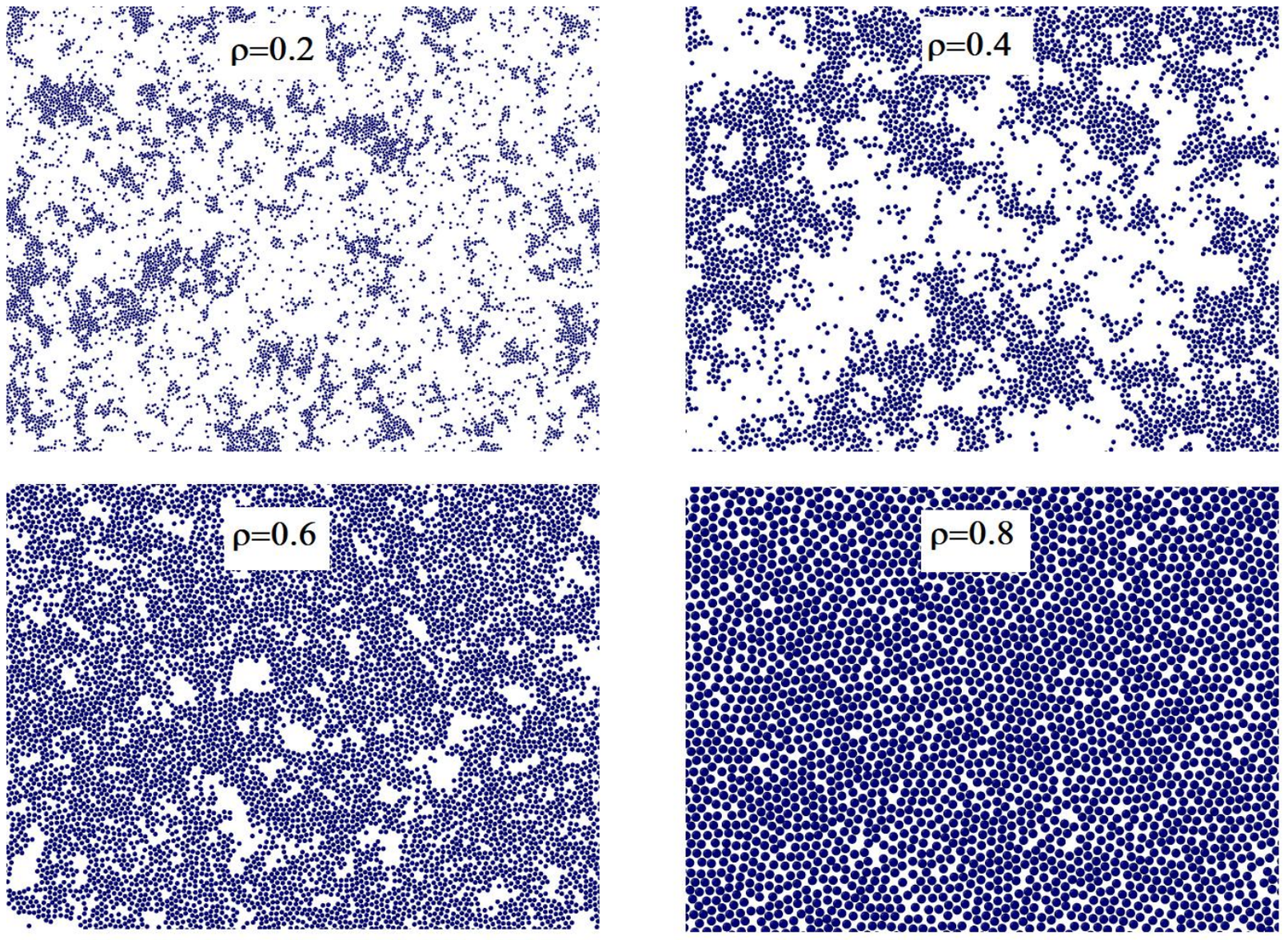}%

\caption{\label{2dlj} Snapshots of 2d LJ system at $T=0.5$ and several densities. Voids are clearly visible at all densities.}
\end{figure}

We show snapshots of the 2d LJ system at a set of densities at the temperature $T=0.5$, which is just above the
critical one in Fig. \ref{2dlj}. It is seen that at low densities the system consists of small clusters and single particles.
When the density increases the size of the clusters becomes larger. After the percolation threshold the system
becomes condensed: the particles belong to several large clusters. However, the voids are clearly visible in the system.
Cavities can be observed even at the density $\rho=0.8$. At the density $\rho=0.9$ the system experiences spontaneous
crystallization. So, one can see that even at the densities close to the freezing one the SCF demonstrates voids.

Another example where voids in SCF were observed in given in Ref. \cite{or-gap}. Molecular dynamics
simulation of carbon with GAP-20 potential \cite{gap-20} was performed in this article. The authors simulated carbon
at $T=5000$, $6000$ and $7000$ K and pressure from 0 to 4 GPa. Voids were found in the system. The authors
calculated the probability distribution of the volumes of the pores and found that their volume
can reach about $10 \AA^3$ for zero pressure. At the same time the authors report the isothermal
compressibility, which decreases with density along isotherms. It looks similar to the case of SCF, where
the compressibility along an isotherm has a maximum at a density near the critical one and then decreases
with density increase.

\begin{figure}

\includegraphics[width=8cm, height=6cm]{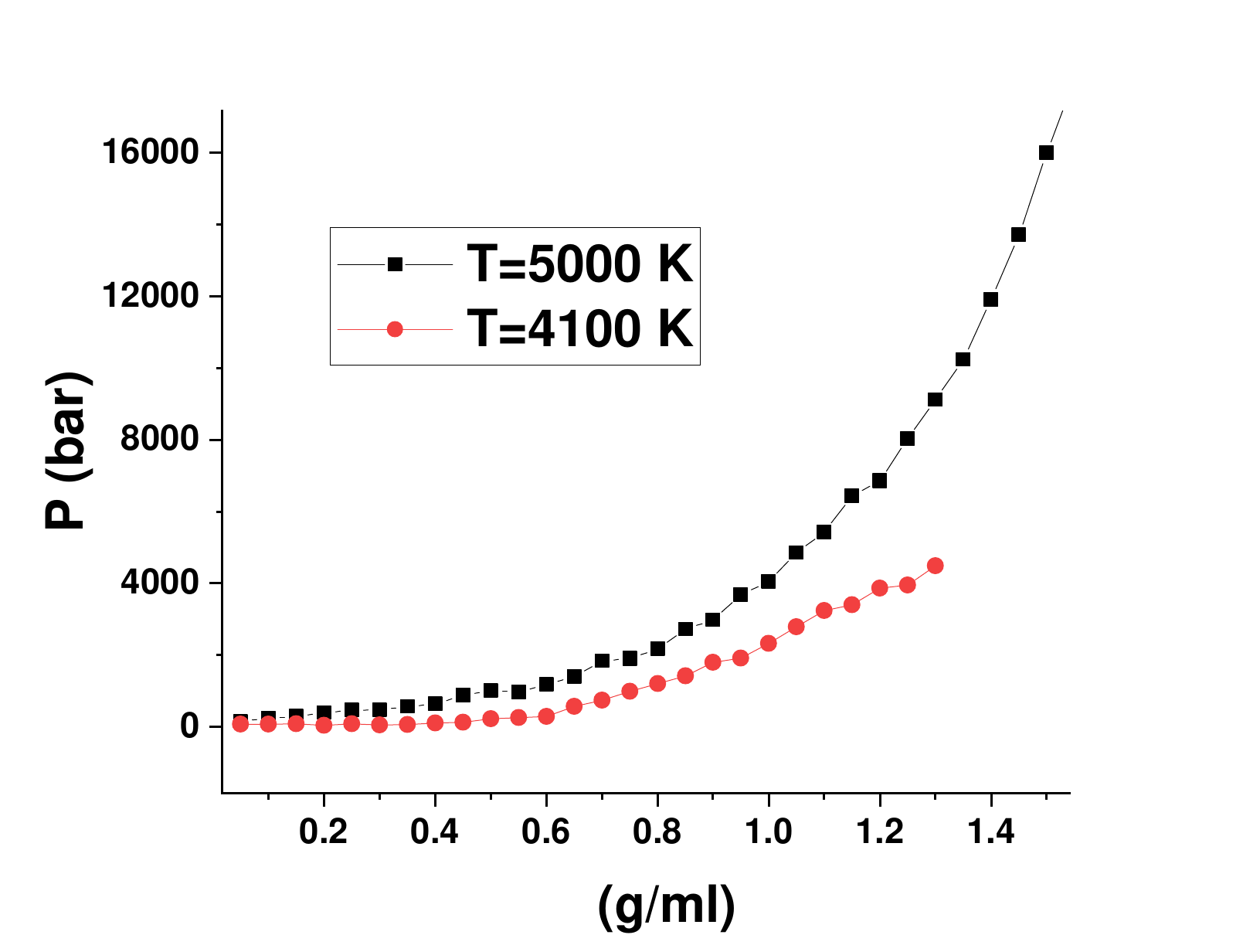}%

\caption{\label{eosc} Equation of state of GAP-20 model of carbon at supercritical ($T=5000$ K) and
subcritical ($T=4100$ K) temperatures.}
\end{figure}

The isotherms of GAP-20 model of carbon were reported in our recent work \cite{carbon-widom}. Here we show equation of
state for two values of temperature: a supercritical and a subcritical ones in Fig. \ref{eosc}. An almost flat
region is seen at $T=4100$ K, which corresponds to a small van der Waals loop, since this  temperature
is close to the critical one. At the same time the pressure monotonously increase with density at $T=5000$ K.
Based on the behavior of the isotherms, the critical point of GAP-20 model of carbon was estimated to be $T=4235$ K \cite{carbon-widom}
$P_c=106.61$ bar and $\rho_c=0.1887$ $g/ml$, i.e. the critical temperature is lower than the ones at which simulation was performed in Ref. \cite{or-gap}.
Therefore, SCF carbon is observed in \cite{or-gap} and this is the reason
for voids to be observed in this study.

\begin{figure}

\includegraphics[width=8cm, height=6cm]{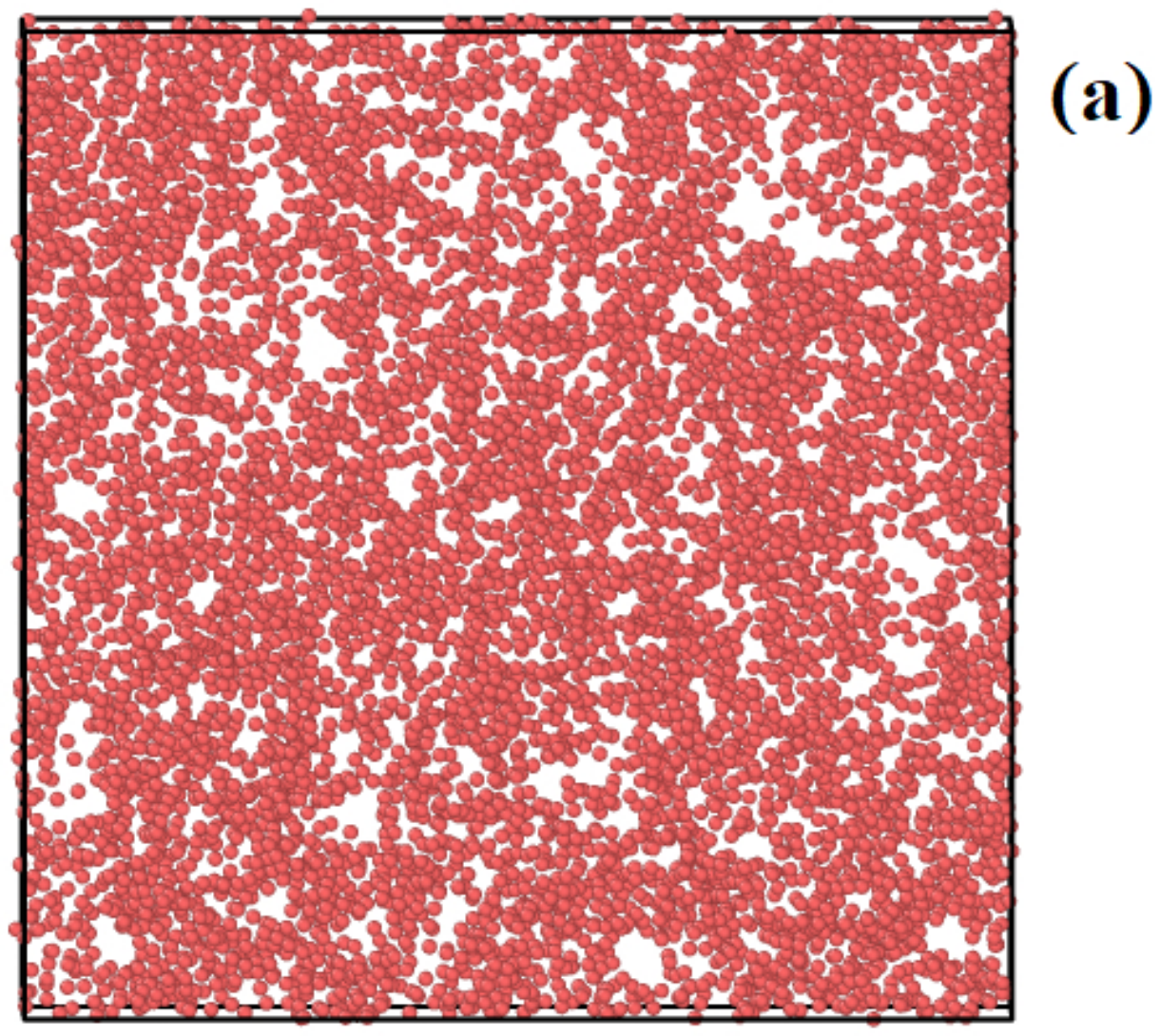}%

\includegraphics[width=9cm, height=7cm]{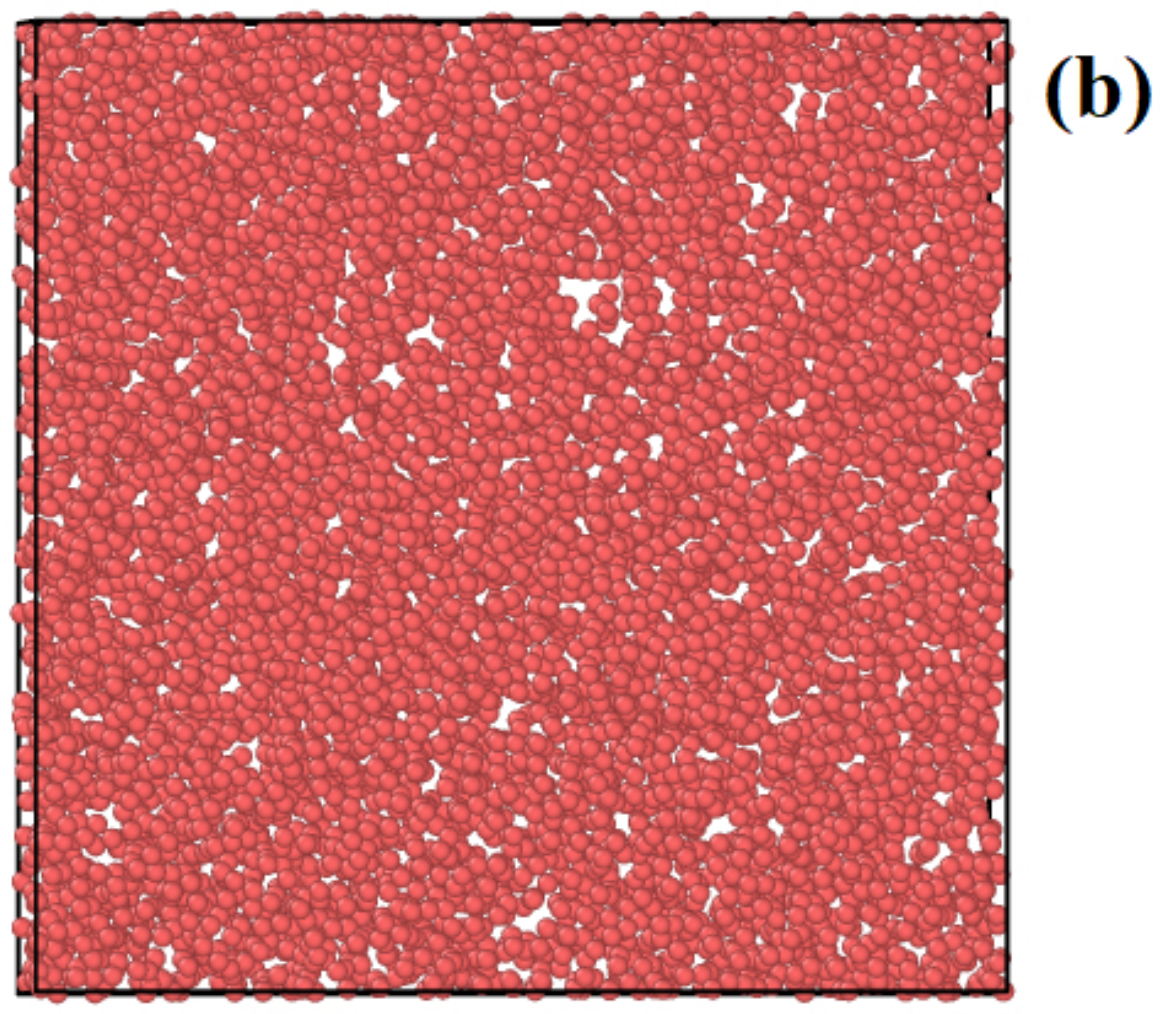}%

\caption{\label{c-sn} Snapshots of GAP-20 model of carbon at $T=5000$ K and (a) $\rho=0.6$ $g/ml$ and (b) $\rho=1.7$ $g/ml$.}
\end{figure}

Figure \ref{c-sn} shows snapshots of GAP-20 model of carbon at $T=5000$ and two values of density: $\rho=0.6$ $g/ml$
(the corresponding pressure $P=1260$ bar) and $\rho=1.7$ $g/ml$ ($P=2.6 \cdot 10^4$ bar). Voids are clearly visible
in both cases, which means that like in 2d LJ system the voids preserve even at very high density: for $\rho=1.7$ $g/ml$
the ratio $\rho/\rho_c=9$. At the same time the voids become smaller with increasing of the density which is apparently
an effect of decreasing of the free volume in the system.

The examples above prove that supercritical fluids demonstrate voids in a wide range of pressures (or densities) up to the melting line.
For this reason, if one observes voids in molecular dynamics simulation of a fluid system, one possibility to be considered
is that the system is above the critical temperature of the gas-liquid phase transition.

\subsection{Liquid-gas two-phase region}

Another case where voids are observed in molecular dynamics simulation of liquid is related to
observation of liquid-gas two phase region. An excellent description of peculiarities of
two-phase regions in molecular dynamics if given in Ref. \cite{prestipino-fs}. In the present
paper we partially repeat the results of this article and give
some extension of the discussion of Ref. \cite{prestipino-fs}.

As it is discussed in Ref. \cite{prestipino-fs}, the periodic boundary conditions make the system to form
drops of liquid of different shape at different average densities. At low density the system forms a spherical
drop. When the density becomes higher, the shape of the drop is changed to be a cylinder. When the density
further increases the particles form a flat slab with a hole. The hole disappears at further density increase.
When the density becomes very close to the liquid branch of the coexistence curve, the system looks like
a condensed phase with a cavity (see Figs. 6 and 7 in \cite{prestipino-fs}). Such changes in the structure
lead also to jumps of the pressure as a function of density along isotherms (Figs. 2-5 of \cite{prestipino-fs}).
These jumps are unphysical and induced by rapid structure changes of the system in periodic boundary conditions.

Typically modern molecular dynamics simulation of liquids involves from several thousands to several dozens
thousands atoms. Although, simulation of much larger systems has been reported \cite{billion,100mil}, the computational burden
of such simulations is very high, and most of researchers do not have enough computational resources.
Typically the density of liquid is several hundred times higher than the one of gas. It allows to say
that the cavity observed in the two-phase region next to the liquid branch of the boiling curve is indeed
a gaseous part of the system: at such system sizes the number of particles in the gas phase is nearly zero.
Therefore, all particles of the system belong to the liquid phase, but from time to time some particles evaporate
into the cavity. After a while they condense in the liquid phase.

\begin{figure*}

\includegraphics[width=16cm, height=14cm]{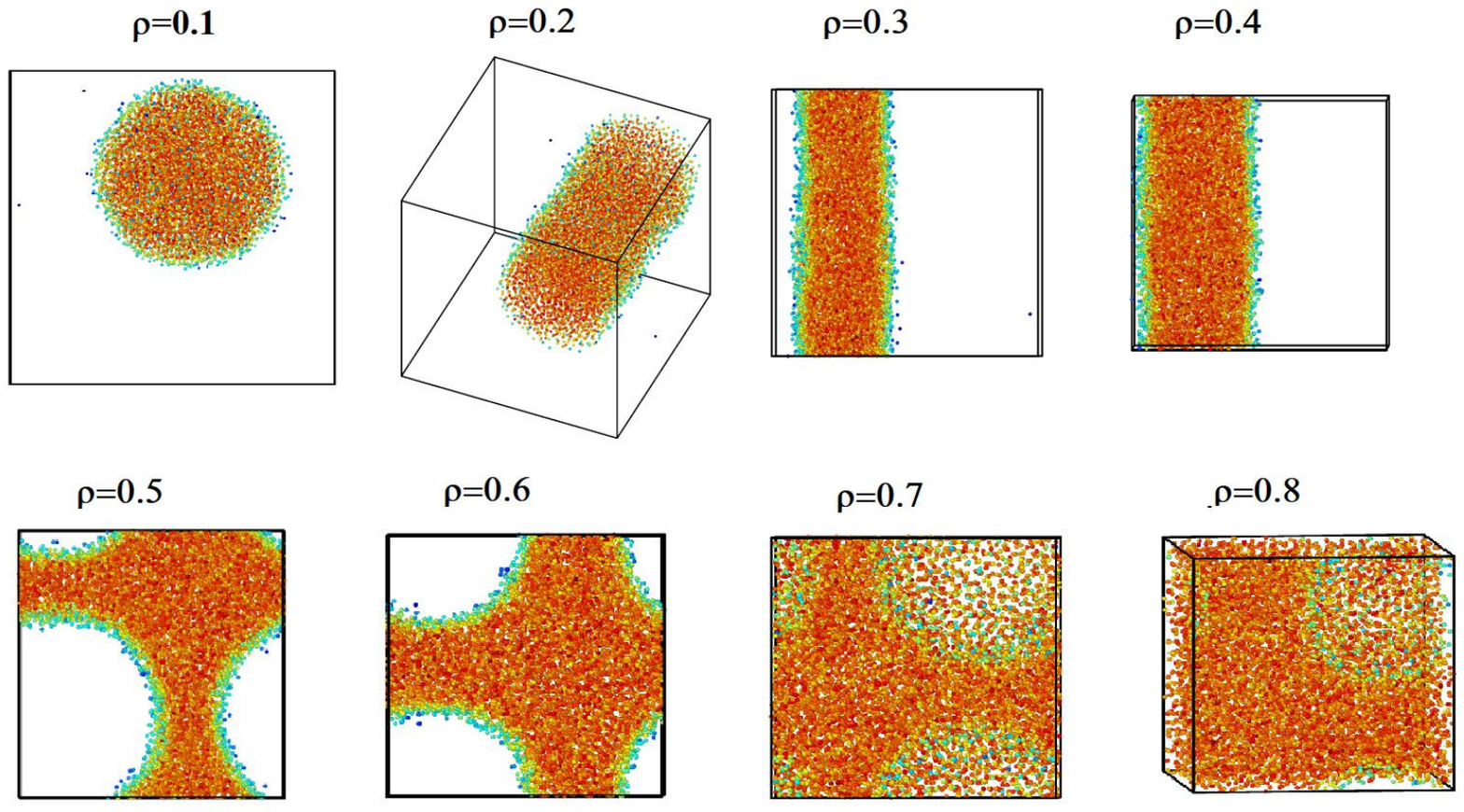}%

\caption{\label{lj05} Snapshots of configurations of 4000 LJ particles at $T=0.5$ and different densities. The particles
are colored by their coordination number from blue (zero nearest neighbors) to red (18 nearest neighbors). The nearest neighbors
are calculated as particles within a sphere of radius $1.6$ around a given particle.}
\end{figure*}

Figure \ref{lj05} shows a set of snapshots of LJ system at $T=0.5$ and different densities. This temperature is
below the temperature of the triple point, therefore coexistence of gas and crystal should be observed. We do not
see the crystallization of the system, but it looks like gas-liquid coexistence. The sequence of snapshots is in agreement
with the one in Ref. \cite{prestipino-fs}: a ball at $\rho=0.1$, a cylinder at $\rho=0.2$, a slab at
$\rho=0.3$ and $0.4$. At $\rho=0.5$ and $0.6$ the system is a liquid with a cylindrical void (periodic boundary conditions
are applied) while at $\rho=0.7$ and $0.8$ the void is spherical. In the case of lower densities one can observe some single
particles in the "vacuum" part of the box, which corresponds to the gaseous phase. If one considers even lower temperature
no particles would be observed in the "vacuum" part which means that the density of saturated vapor is very low at this
temperature.

\begin{figure*}

\includegraphics[width=16cm, height=14cm]{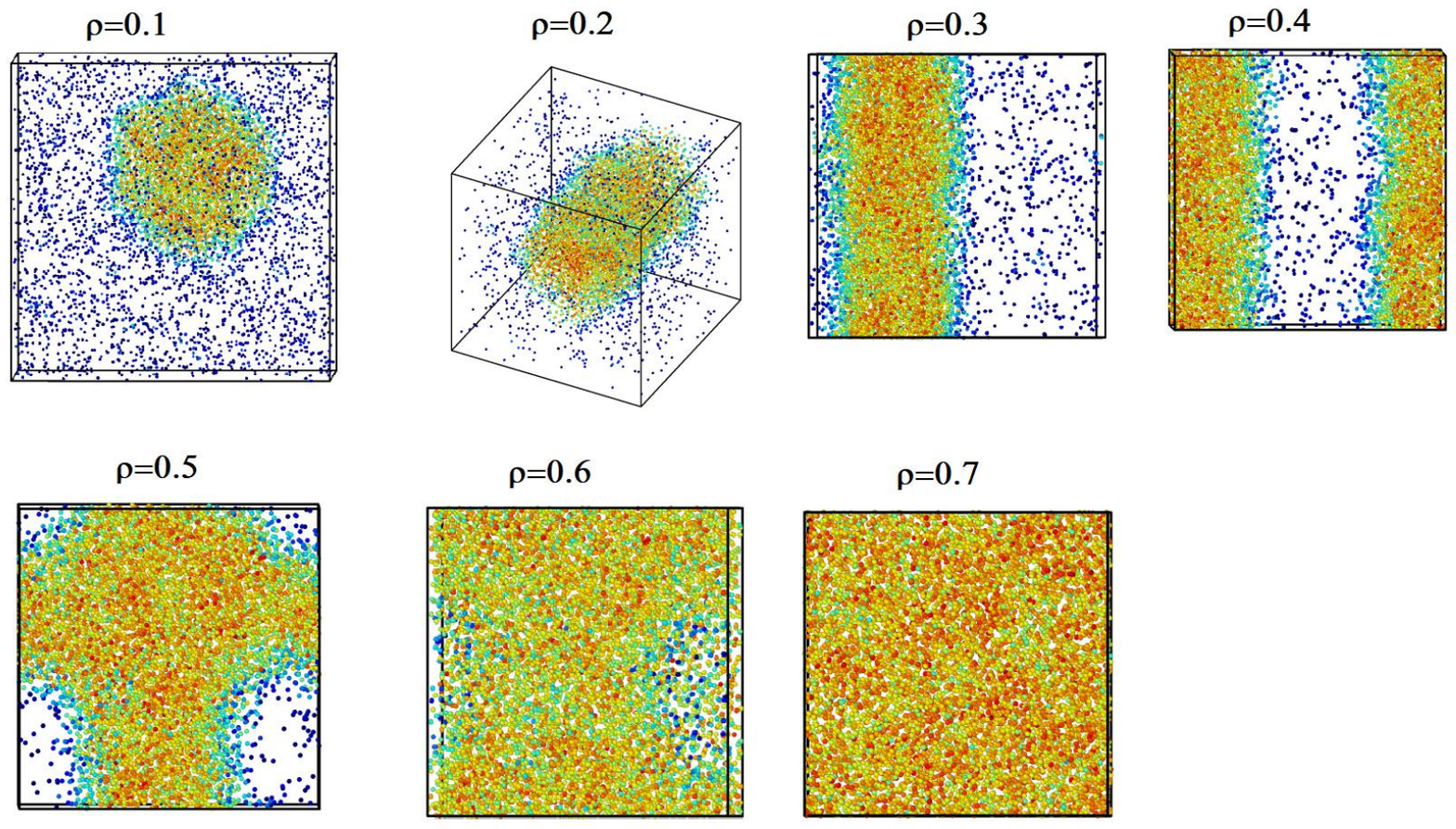}%

\caption{\label{lj1} Snapshots of configurations of 4000 LJ particles at $T=1.0$ and different densities. The particles
are colored by their coordination number from blue (zero nearest neighbors) to red (17 nearest neighbors). The nearest neighbors
are calculated as particles within a sphere of radius $1.6$ around a given particle.}
\end{figure*}

A set of snapshots of the LJ system of 8000 particles at $T=1.0$ is shown in Fig. \ref{lj1}.
Qualitatively the results are similar to the ones at $T=0.5$:
a spherical drop of liquid in the gas at $\rho=0.5$, a cylindrical drop at $\rho=0.2$, a slab of liquid at $\rho=0.3$ and
$0.4$, a cylindrical void at $\rho=0.5$ and a spherical void at $\rho=0.6$. At the density $\rho=0.7$ there are
just small voids in the bulk of liquid phase. However, at this temperature the number of particles in the gaseous
phase is substantial. As a result a coexistence of liquid and gas is observed in the system.

\begin{figure*}

\includegraphics[width=16cm, height=14cm]{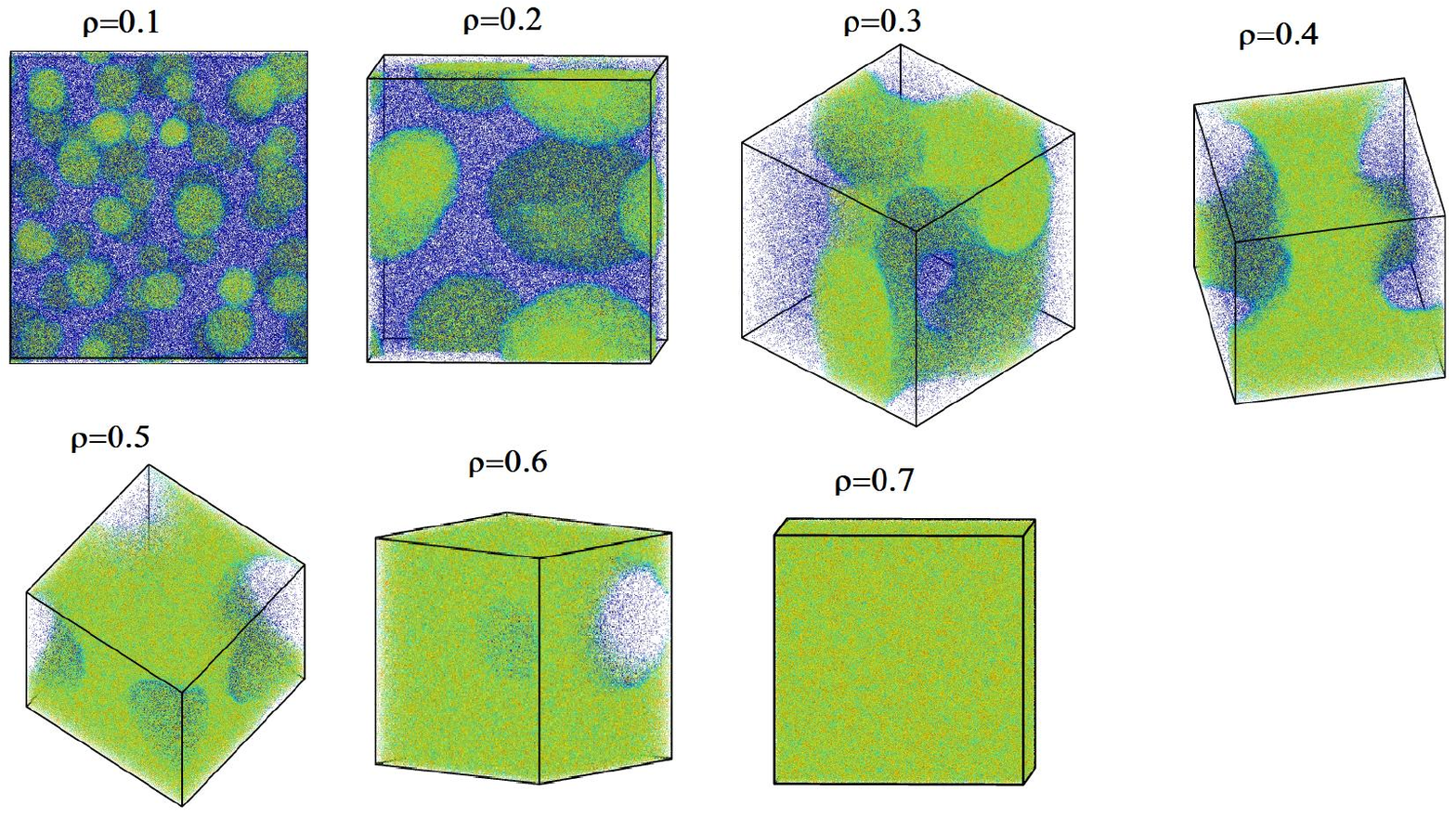}%

\caption{\label{lj-mil} Snapshots of configurations of $10^6$ LJ particles at $T=1.0$ and different densities. The particles
are colored by their coordination number from blue (zero nearest neighbors) to red (20 nearest neighbors). The nearest neighbors
are calculated as particles within a sphere of radius $1.6$ around a given particle.}
\end{figure*}

In Figure \ref{lj-mil} we show snapshots of the LJ system of a million particles at $T=1.0$ and a set of densities. Qualitatively
the system behaves like the one of 8000 particles, but some observed 'structures' have changed. In the case of $\rho=0.1$
the system consist of a set of liquid balls inside the vapor phase. In the case of $\rho=0.2$ the liquid drops become
ellipses (cylinder in the smaller system). The aspect ratio of the cylinders is rather small, so one can assume that
this is spherical drops with strong shape fluctuations. At $\rho=0.3$ the liquid part of the system has some
irregular shape, which consists of condensed phase with voids.
Similar irregular shape structures are observed in molecular simulation of phase separation
of diblock copolymers \cite{polymer}. The structure becomes simpler at higher densities: at $\rho=0.4$: it is again
a liquid phase with a cylindrical void and at $\rho=0.5$ and $0.6$ the void becomes spherical. The system looks
like a condensed liquid phase at $\rho=0.7$.

Comparing the results for two system sizes we see that the qualitative behavior is the same, but the set of structures in the big
system is slightly different from the one of the small one. This is not surprising. It is well known that in the thermodynamic limit
a two phase liquid-gas system should be a spherical drop of liquid surrounded by the vapor (no gravity is assumed). The complex
structures observed in simulations are the finite size effects and the effects of the periodic boundary conditions. For this reason
it is not surprising that these effects are also influenced by the size of the system.

In the present paper we demonstrate the finite size effects on the structure of a two phase system with the LJ fluid.
However, these effects are general and can be observed in any kind of fluid systems. To demonstrate it we
show snapshots of the two-phase region at different densities of a TIP4P/2005 model of water in Supplementary materials.


\subsection{Difference between NVT and NPT ensembles in two-phase region}

\begin{figure}

\includegraphics[width=9cm, height=8cm]{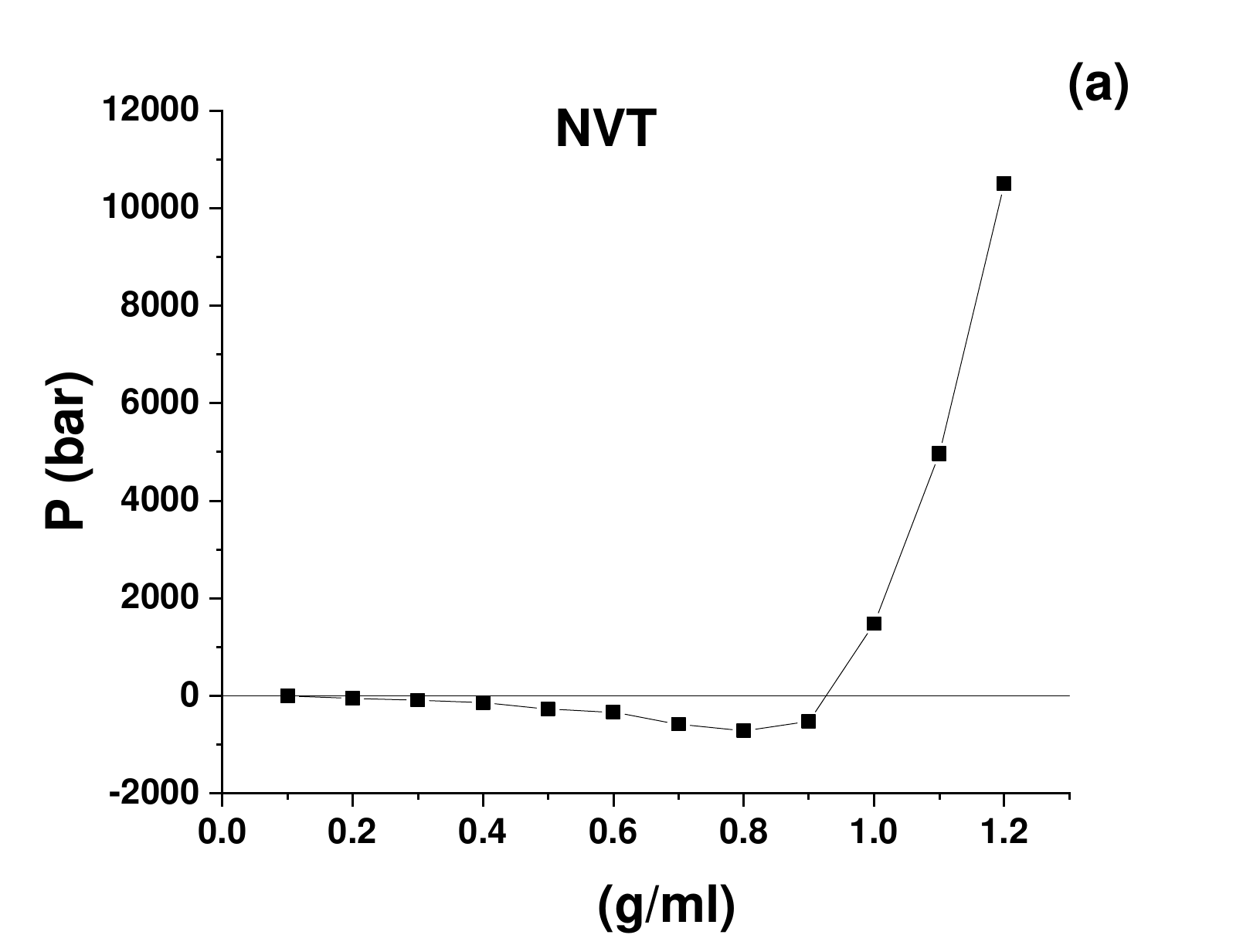}%

\includegraphics[width=9cm, height=8cm]{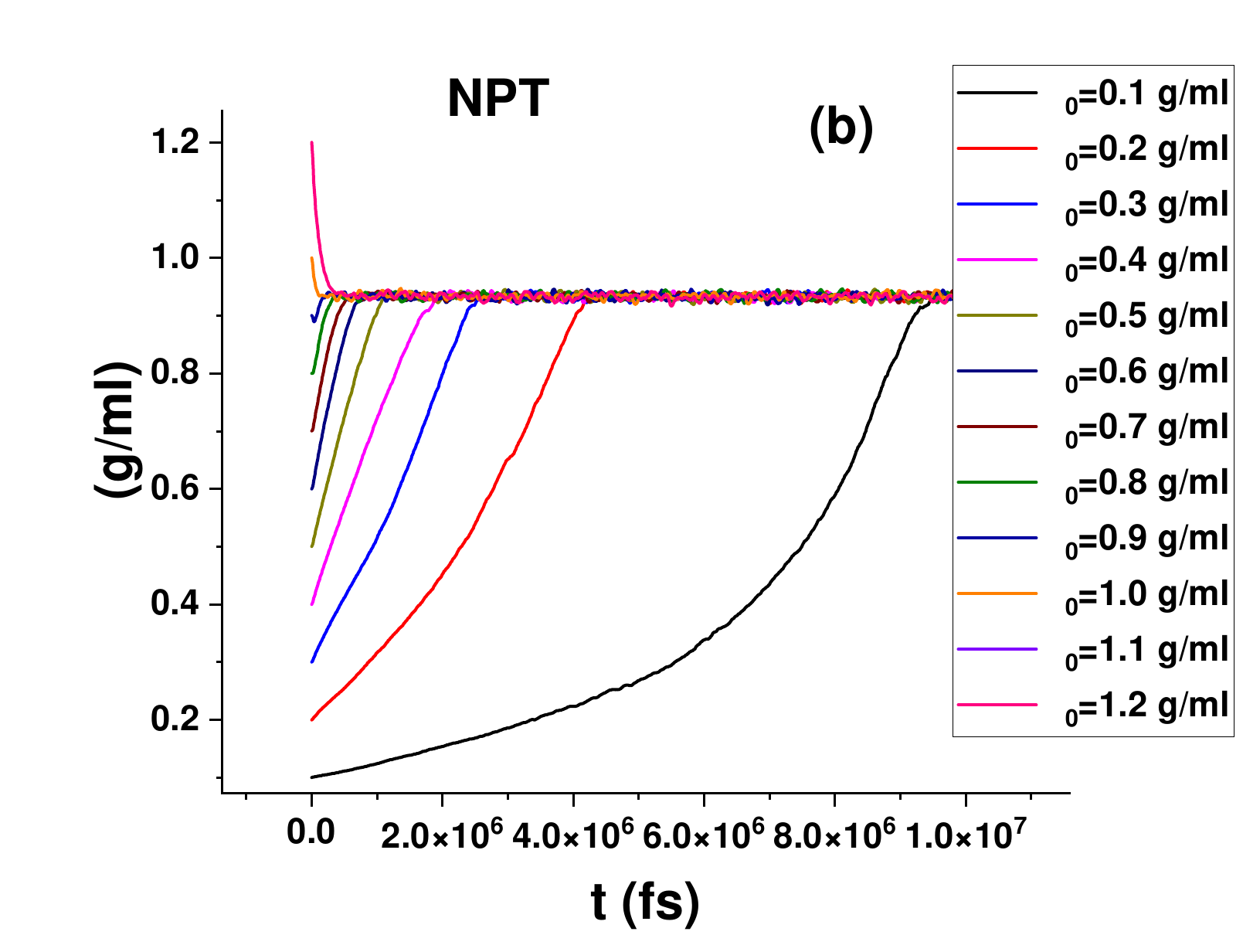}%

\caption{\label{nvtnpt} Comparison of simulation of water in (a) NVT and (b) NPT ensembles. $T=400$ K.}
\end{figure}

In the case of two-phase region simulation in NVT and NPT lead to different results. We have discussed the structure of
a fluid simulated in NVT ensemble at the conditions of two-phase region above. Equation of state of
water at $T=400$ K obtained in simulation in canonical ensemble if shown in Fig. \ref{nvtnpt} (a). It is seen that in this case
a van der Waals loop appears which is a trace of the first-order phase transition. The pressure becomes negative in the
coexistence region. The location of the transition points can be found by Maxwell construction (the EoS should
be redrawn as pressure as a function of volume to perform the Maxwell construction).

In the case of NPT ensemble the density of the system rapidly changes and converges to the value close to the
one at the coexistence line. In Fig. \ref{nvtnpt} (b) we show the evolution of the density of the system
starting from different initial values. The initial density varied from $0.1$ to $1.2$ $g/ml$ with the step
$\Delta \rho=0.1$ $g/ml$. In all cases the system converged to the final density $\rho_f=0.93$ $g/ml$,
which is within the error bar at the coexistence point \cite{vega-vle}.


This feature of the NPT simulation of two-phase regions is described in the well recognized text book "Understanding
molecular simulation" by D. Frenkel and B. Smit \cite{book-fs}. The authors give an example of a simple and rapid method
to find an approximate location of the density of liquid at the boiling curve as a result of NPT simulation
at zero pressure. Our results confirm the applicability of this rough estimation. The authors state that the
resulting density is only a rough estimate of the liquid density at the coexistence curve. In our point
of view the main origin of the error is zero pressure. It is implicitly assumed that if the temperature is just above the
triple point the pressure is very low, i.e., nearly zero. If the simulations are performed at relatively low temperature,
this assumption is reasonable. However, it becomes not valid at higher temperature, especially in the vicinity
of the critical point.

\begin{figure}

\includegraphics[width=9cm, height=8cm]{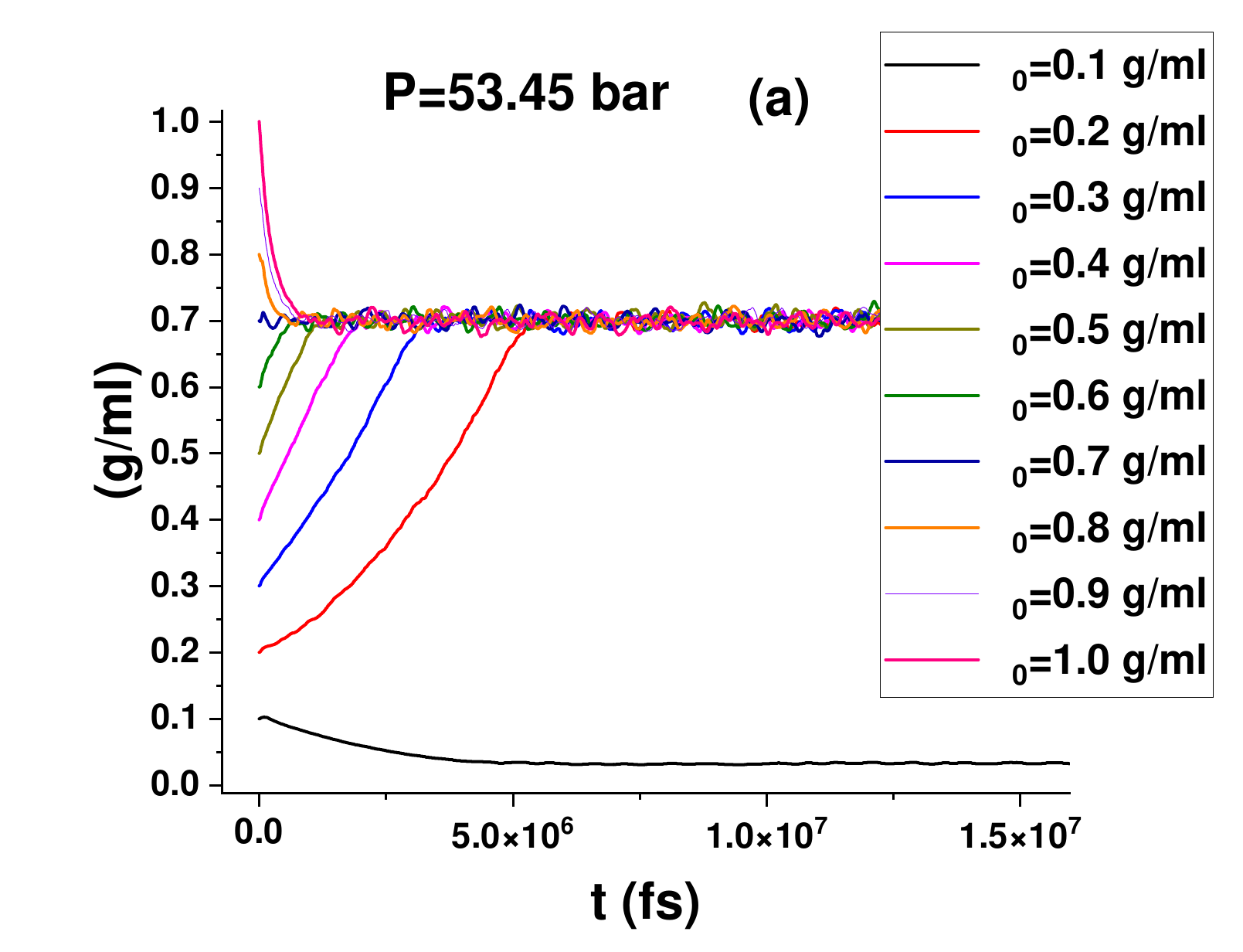}%

\includegraphics[width=9cm, height=8cm]{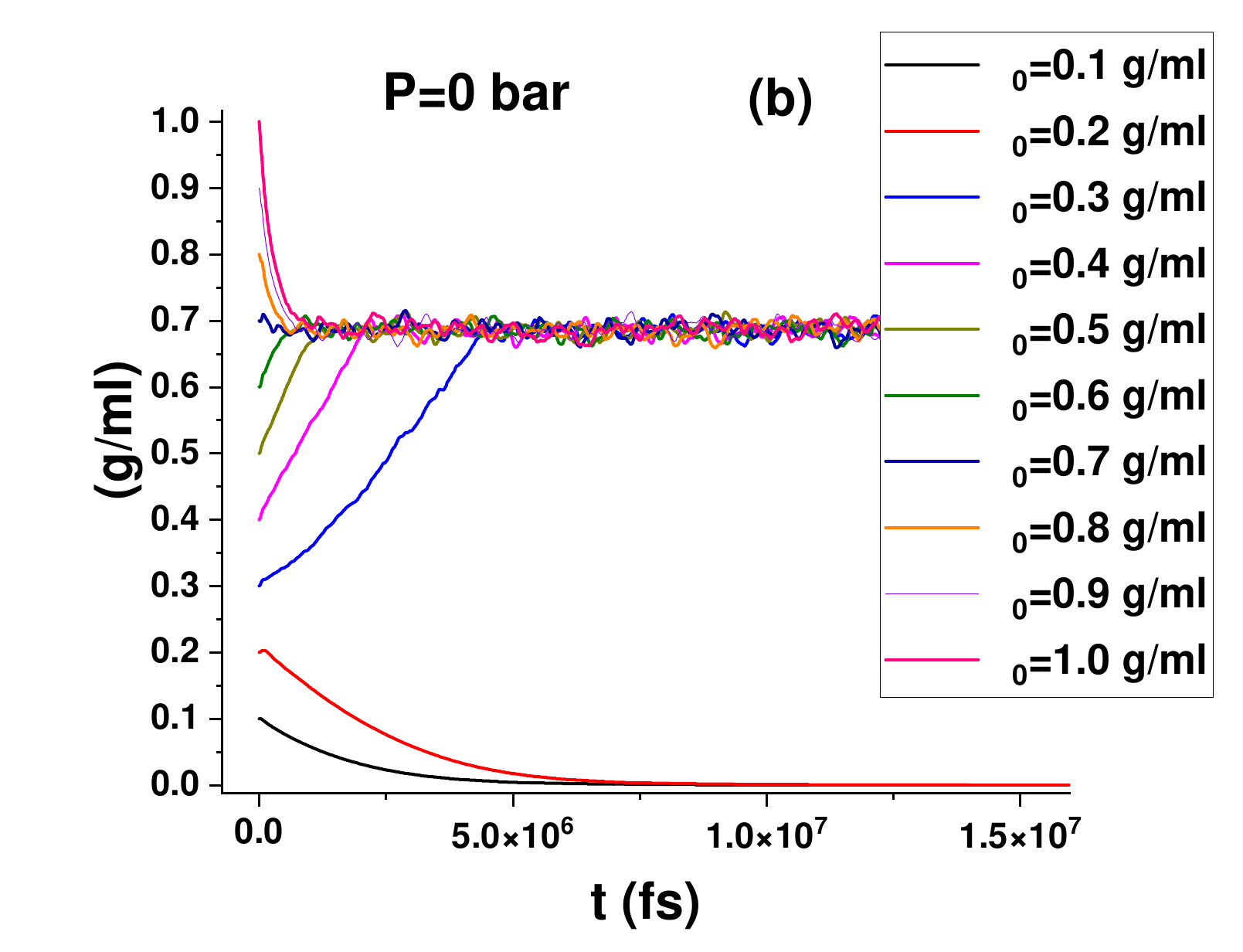}%

\includegraphics[width=9cm, height=8cm]{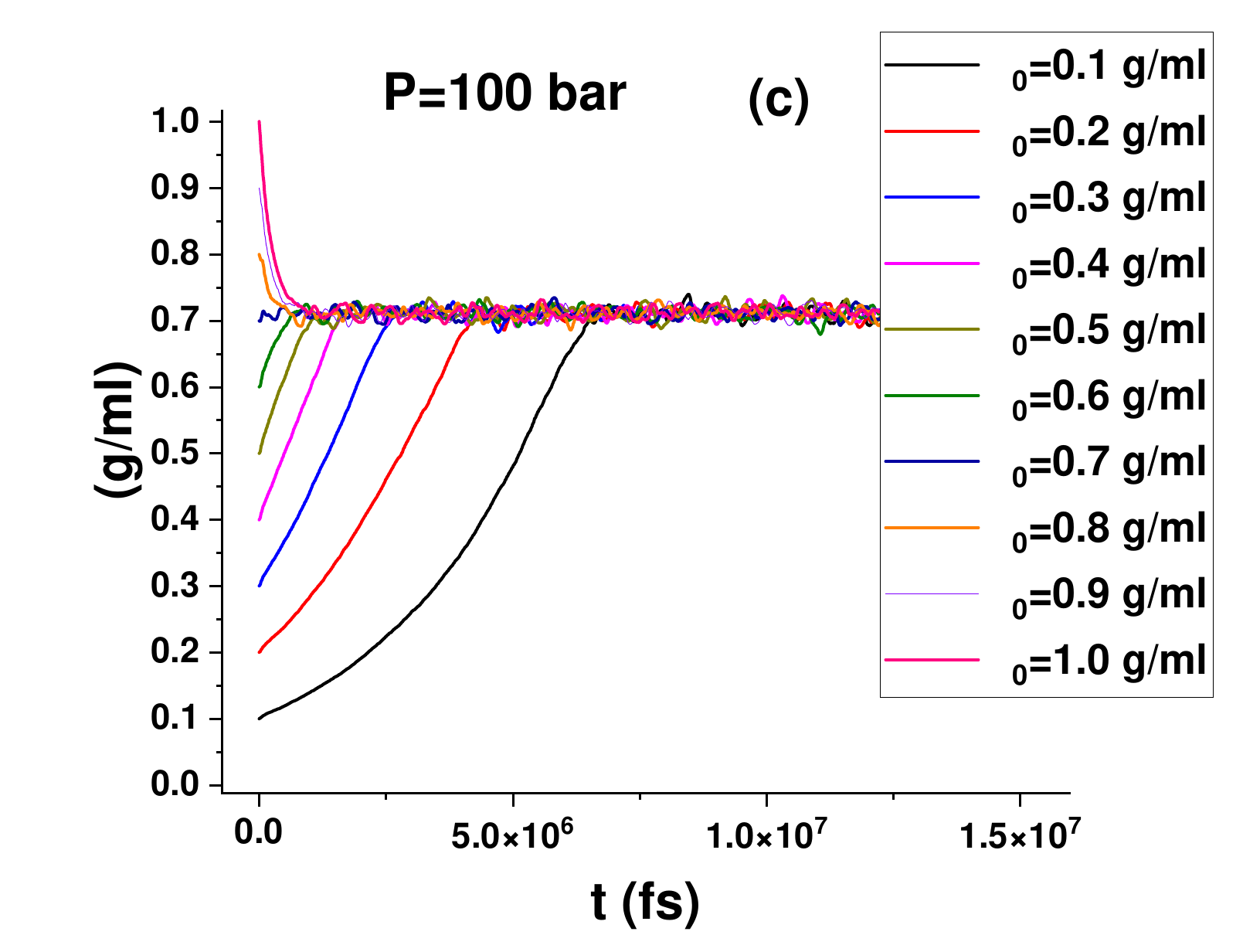}%

\caption{\label{wat570} Relaxation of density of water starting from different initial ones at $T=570$ K and different pressures.
(a) $P=53.45$ bar (coexistence); (b) $P=0$ bar and (c) $P=100$ bar.}
\end{figure}

In order to check the applicability of the method at finite pressure we perform simulation of water at $T=570$ K
and $P=53.45$ bar, which is the coexistence pressure at this temperature \cite{vega-vle}. The results are given in
Fig. \ref{wat570} (a). It is seen that in the case of low initial density $\rho_0=0.1$ $g/ml$ the volume of the
system enlarges in the course of simulation and the system becomes gaseous. However, if the initial density is
$\rho_0=0.2$ $g/ml$ or larger the volume of the system decreases and the system evolves into the liquid phase.
The density of the liquid is $\rho_l=0.701$ $g/ml$ which is very close to the one obtained in the framework of
Gibbs-Duhem integration scheme \cite{vega-vle}. The density of gas obtained in the present study is $\rho_g=0.033$ $g/ml$
which is also in good agreement with the Gibbs-Duhem results: $\rho_g=0.0316$ $g/ml$ \cite{vega-vle}. Importantly, when
the system comes into the gaseous phase the box strongly fluctuates, which is consistent with low value of
compressibility coefficient of the gas. For this reason, larger error in equilibrium density of gas is expected.

Figure \ref{wat570} (b) shows simulation of the same system at zero pressure. Like in the previous case the system
goes into the gas phase if the initial density is small and to the liquid one if the initial density is large enough.
The final values of the density is $\rho_g=1.3 \cdot 10^{-4}$ $g/ml$ for the gas and $\rho_l=0.69$ $g/ml$ for the liquid.
This result is in good agreement with the assertion of the Frenkel and Smith book that zero pressure simulation gives
reasonable, but not exact values of the density of liquid at the coexistence.

In the case of pressure higher than the one at coexistence the volume of the system decreases even if the initial
density is low (Fig. \ref{wat570} (c)), therefore the system relaxes into the liquid state. The final density is $\rho_l=0.712$
$g/ml$, which is larger than the density at coexistence.

These results show that the simulation at constant volume and at constant pressure lead to different result, when they
are performed in the liquid-gas two-phase region. The densities of coexisting phases can be precisely determined in NPT
simulation, which, however, requires the knowledge of the equilibrium pressure at given temperature.

\subsection{A case of a complex liquid: tellurium}

The described feature of NVT and NPT ensembles allows us to find out that the system is inside the two phase
region even in the case of very small system, where direct observation of the coexisting phases is difficult.
Examples of such systems can be found in ab-initio simulations of liquids.

As it was discussed above, liquid tellurium demonstrates numerous anomalous features. In particular, it was
found within the ab-initio simulation methods, that a large part of volume of liquid tellurium belongs to
cavities \cite{te}. Temperatures from $560$ to $970$ K were simulated in this work. The melting point of
tellurium is $T_m=722.6$ K and the boiling point of tellurium is $T_b=1260$ K. Therefore the temperatures
of the Ref. \cite{te} are partially in liquid and partially in supercooled state.

Another first principles simulation of liquid tellurium is reported in Ref. \cite{te-1}. The authors
calculate the pressure as a function of density at $T=1123$ K for both LDA and GGA exchange-correlation
functionals in NVT ensemble. Interestingly, at the densities $rho< 6.0$ $g/ml$ both functionals give negative pressure,
which means that the system is within two phase region at these conditions. It means that DFT calculations
underestimate the boiling point of Te.

In the case of Ref. \cite{te} the calculations are also performed constant density and the
values of the density are taken from experiment. However, as is seen from the discussion above,
the ab-initio simulation can underestimate the boiling point, and the experimental density can
appear in the two-phase region when simulation is performed.

In order to check this assumption we perform NPT simulation of liquid tellurium at several temperatures.
The density of liquid tellurium at zero pressure and $T=934$ K is $\rho=6.03$ $g/ml$ and at $T=1028$ K
the density is $\rho=5.98$ $g/ml$, which is higher than the results of Ref. \cite{te} ($\rho=5.62$ $g/ml$ at $T=970$ K).
It means that the results of Ref. \cite{te} correspond to liquid-gas two phase region. These results are also consistent
with negative pressures obtained in Ref. \cite{te} for the temperatures $722$ and $970$ K (see Table II of Ref. \cite{te}).
At the same time, liquid tellurium does demonstrate rings and chains, like it was stated in Ref. \cite{te}. Moreover,
liquid tellurium is a very complex system, which demonstrates numerous anomalies, like the density anomaly (negative value
of thermal expansion coefficient) \cite{te-den,te-anom}, anomalous specific heat \cite{te-anom,te-cp}, speed of
sound \cite{te-cs1,te-cs2}, etc. An possible explanation of these anomalies was proposed in Ref. \cite{br-anom}:
it can be related to a smeared phase transformation of liquid tellurium.

We would like also to mention that liquid tellurium is a very complex system, which is very difficult to simulate.
Liquid tellurium is a metal, which means than strong ion-electron correlation take place \cite{sim-met},
difficulties with convergence of self consistent calculation. As it was discussed above, the thermodynamic
conditions of liquid tellurium are close to boiling line which requires proper accounting of van der Waals
correction \cite{sim-wat}, which was not employed in Refs. \cite{te} and \cite{te-1}. We did not take into
account the van der Waals corrections, since our goal was to reproduce the results of previous calculations
and show the problems of interpretation of these calculations. Although the calculations of Refs. \cite{te,te-1} were
performed at the highest level of the time when they were made, it looks clear now, that more sophisticated
computational schemes are required for liquid tellurium, which goes beyond the scope of the present paper.

\section{Conclusions}

In the present paper we discuss the origin of large voids observed in molecular simulations of "liquids". We show
that large voids appear in two cases: either the system is above the critical temperature of gas-liquid transition
or it is within the liquid-gas two phase region. In the case of two-phase region the NVT and NPT simulations
give different results. While in the NVT ensemble negative pressure is observed, the NPT ensemble tends
to bring the system into a single phase region: gas, if the starting density is very low, or liquid,
if the starting density is high. Several examples of system with large voids are discussed. In particular,
it is shown that voids in liquid tellurium comes from misinterpretation of ab-initio molecular dynamics data.

\section{Acknowledgments}

This work has been carried out using computing resources of the federal
collective usage center Complex for Simulation and Data Processing for
Mega-science Facilities at NRC "Kurchatov Institute", http://ckp.nrcki.ru/.
The work was supported by the Russian Science Foundation (Grant 25-22-00876).


\end{document}